\documentstyle[preprint,aps, eqsecnum, epsf]{revtex}
\tightenlines  

\def\epsfig#1#2#3#4
         {
         \epsfysize=#2 \vbox{ \hglue#3 \epsfbox[#4]{#1} }
         }
\def\epsfigrot#1#2#3#4
         {
         \epsfxsize=#2
         \setbox\rotbox=\hbox to #2{\epsfbox[#4]{#1}}
         \vbox{\hglue#3 \rotl\rotbox}
         }
\newbox\rotbox

\begin{document}
 
\title{
Quasi-Spin-Charge  Separation and the Spin Quantum Hall Effect
}
\author{D.Bernard$^{1}$ 
 \footnote{Member of the C.N.R.S.} and  A. LeClair$^{1,2}$}
\address{ ${}^1$ Service de Physique 
Theorique de Saclay, F-91191, Gif-sur-Yvette, France}
\address{  ${}^2$ Newman Laboratory, Cornell University, Ithaca, NY
14853.}
\address{}
\date{February 29, 2000}
\maketitle

\begin{abstract}

We use quantum field theory methods to study the network model for the
spin  quantum Hall transition.  When the couplings are fine tuned in a certain way,
the spin and charge degrees of freedom,  corresponding to the supercurrent algebras
$su(2)_0 $ and $osp(2|2)_{-2}$ respectively, decouple in the renormalization
group flow.  
The infrared fixed point of this simpler theory is the coset
$osp(4|4)_1 / su(2)_0$ which is closely related to the current
algebra $osp(2|2)_{-2}$ but not identical. 
 Some  critical exponents are computed
and shown to agree with the recent predictions based on percolation.

\end{abstract}
\vskip 0.2cm
\pacs{PACS numbers: }
\narrowtext
%
%
%
\def\oti{{\otimes}}
\def\bra#1{{\langle #1 |  }}
\def\lb{ \left[ }
\def\rb{ \right]  }
\def\tilde{\widetilde}
\def\bar{\overline}
\def\hat{\widehat}
\def\*{\star}
\def\[{\left[}
\def\]{\right]}
\def\({\left(}		\def\BL{\Bigr(}
\def\){\right)}		\def\BR{\Bigr)}
%
%
\def\zb{{\bar{z} }}
\def\zbar{{\bar{z} }}
\def\frac#1#2{{#1 \over #2}}
\def\inv#1{{1 \over #1}}
\def\half{{1 \over 2}}
\def\d{\partial}
\def\der#1{{\partial \over \partial #1}}
\def\dd#1#2{{\partial #1 \over \partial #2}}
\def\vev#1{\langle #1 \rangle}
\def\ket#1{ | #1 \rangle}
\def\rvac{\hbox{$\vert 0\rangle$}}
\def\lvac{\hbox{$\langle 0 \vert $}}
\def\2pi{\hbox{$2\pi i$}}
\def\e#1{{\rm e}^{^{\textstyle #1}}}
\def\grad#1{\,\nabla\!_{{#1}}\,}
\def\dsl{\raise.15ex\hbox{/}\kern-.57em\partial}
\def\Dsl{\,\raise.15ex\hbox{/}\mkern-.13.5mu D}
%
%
\def\th{\theta}		\def\Th{\Theta}
\def\ga{\gamma}		\def\Ga{\Gamma}
\def\be{\beta}
\def\al{\alpha}
\def\ep{\epsilon}
\def\vep{\varepsilon}
\def\la{\lambda}	\def\La{\Lambda}
\def\de{\delta}		\def\De{\Delta}
\def\om{\omega}		\def\Om{\Omega}
\def\sig{\sigma}	\def\Sig{\Sigma}
\def\vphi{\varphi}
%
%
\def\CA{{\cal A}}	\def\CB{{\cal B}}	\def\CC{{\cal C}}
\def\CD{{\cal D}}	\def\CE{{\cal E}}	\def\CF{{\cal F}}
\def\CG{{\cal G}}	\def\CH{{\cal H}}	\def\CI{{\cal J}}
\def\CJ{{\cal J}}	\def\CK{{\cal K}}	\def\CL{{\cal L}}
\def\CM{{\cal M}}	\def\CN{{\cal N}}	\def\CO{{\cal O}}
\def\CP{{\cal P}}	\def\CQ{{\cal Q}}	\def\CR{{\cal R}}
\def\CS{{\cal S}}	\def\CT{{\cal T}}	\def\CU{{\cal U}}
\def\CV{{\cal V}}	\def\CW{{\cal W}}	\def\CX{{\cal X}}
\def\CY{{\cal Y}}	\def\CZ{{\cal Z}}

\def\rvac{\hbox{$\vert 0\rangle$}}
\def\lvac{\hbox{$\langle 0 \vert $}}
\def\comm#1#2{ \BBL\ #1\ ,\ #2 \BBR }
\def\2pi{\hbox{$2\pi i$}}
\def\e#1{{\rm e}^{^{\textstyle #1}}}
\def\grad#1{\,\nabla\!_{{#1}}\,}
\def\dsl{\raise.15ex\hbox{/}\kern-.57em\partial}
\def\Dsl{\,\raise.15ex\hbox{/}\mkern-.13.5mu D}
\def\Asl{\,\raise.15ex\hbox{/}\mkern-.13.5mu A}
%
%
%
\font\numbers=cmss12
\font\upright=cmu10 scaled\magstep1
\def\stroke{\vrule height8pt width0.4pt depth-0.1pt}
\def\topfleck{\vrule height8pt width0.5pt depth-5.9pt}
\def\botfleck{\vrule height2pt width0.5pt depth0.1pt}
\def\Zmath{\vcenter{\hbox{\numbers\rlap{\rlap{Z}\kern
0.8pt\topfleck}\kern 2.2pt
                   \rlap Z\kern 6pt\botfleck\kern 1pt}}}
\def\Qmath{\vcenter{\hbox{\upright\rlap{\rlap{Q}\kern
                   3.8pt\stroke}\phantom{Q}}}}
\def\Nmath{\vcenter{\hbox{\upright\rlap{I}\kern 1.7pt N}}}
\def\Cmath{\vcenter{\hbox{\upright\rlap{\rlap{C}\kern
                   3.8pt\stroke}\phantom{C}}}}
\def\Rmath{\vcenter{\hbox{\upright\rlap{I}\kern 1.7pt R}}}
\def\Z{\ifmmode\Zmath\else$\Zmath$\fi}
\def\Q{\ifmmode\Qmath\else$\Qmath$\fi}
\def\N{\ifmmode\Nmath\else$\Nmath$\fi}
\def\C{\ifmmode\Cmath\else$\Cmath$\fi}
\def\R{\ifmmode\Rmath\else$\Rmath$\fi}
%
%
\def\debut{ \begin{eqnarray} }
\def\fin{ \end{eqnarray} }
\def\non{ \nonumber }
%








\def\beq{\begin{equation}}
\def\eeq{\end{equation}}

\section{Introduction}

Recently a number of works have appeared investigating a new universality
class of delocalization transition referred to as the 
spin quantum Hall effect.  This transition can occur in certain
dirty superconductors with unbroken $su(2)$ spin-rotation symmetry.  
Kagalovsky et. al. constructed an $su(2)$ invariant network model for
the transition and numerically determined some critical
exponents\cite{Kagalovsky}.  Senthil et. al. on the other hand modeled the 
phenomena with a supersymmetric spin chain\cite{Senthil1,Senthil}.  
Remarkably, the network model  was mapped directly on the lattice 
onto classical percolation 
by Gruzberg et. al. and exact exponents were computed which agreed very well
with the numerical simulations of the super spin chain\cite{Read}.  

In the quantum field theory approach to delocalization transitions, the computation 
of critical exponents is normally a difficult strong-coupling problem requiring 
the existence of non-trivial infrared fixed points.  The claim that for the spin
quantum Hall effect the infrared fixed point of the disorder averaged effective
field theory is simply percolation is rather unexpected and for this reason 
we set out to understand this using quantum field theory methods.  
Our starting point is the hamiltonian formulation of the network model given in
\cite{Kagalovsky}.  We carry out the  disorder averaging  using the supersymmetric
method in conjunction with conformal field theory methods. 
This leads to an effective action which consists of a 
conformal field theory with an $osp(4|4)$ super-current algebra symmetry 
perturbed by certain marginal operators which are bilinear in the supercurrents.

The one-loop
renormalization group (RG) $\beta$eta functions we compute for the three independent
couplings appear to be as complicated as those for the usual quantum Hall 
transition\cite{carg}.   There exists  
a fine-tuning of the network model couplings wherein two of the couplings are
essentially identified and the resulting model  is remarkably simpler and 
can be solved. This is due in part to the fact
that the stress tensor for the $osp(4|4)$ current algebra can be written
as the sum of the stress tensors for the  
$osp(2|2)$ level $-2$ and $su(2)$ level $0$ current
algebras corresponding to the charge and spin degrees of freedom respectively. 
This ``spin-charge separation'' is present in the disorder-averaged effective
theory and this leads to two decoupled $\beta$eta functions.  
This simplification allows us to determine the 
non-trivial infrared fixed point:  the $su(2)$ level $0$ degrees of freedom
decouple in the flow and the infrared conformal field theory is simply
the coset $osp(4|4)_1 /su(2)_0$. 
In this way we recover some of the percolation 
exponents predicted in \cite{Read}.  

Due to the logarithmic nature of the above conformal field theories, 
we find that in spite of  the separation of the stress tensor into commuting
pieces,  $T_{osp(4|4)} = T_{osp(2|2)} + T_{su(2)} $,  the Hilbert space
does not factorize.  This leads to the peculiar result that 
the coset 
$osp(4|4)_1 / su(2)_0 $ is {\it not} equivalent to the $osp(2|2)_{-2}$ current
algebra theory,   
even though this coset possesses the current algebra as a symmetry and
the conformal dimensions of the coset are the same as for the
$osp(2|2)_{-2}$ current algebra.

In the last section we study the possible universality classes
based on the 1-loop RG equations and suggest that the network model 
is universally attracted  to a so-called ``strange direction''.

\section{The Models} 

\def\sigvec{\vec{\sigma}}
\def\alvec{\vec{\alpha}}

Kagalovsky et. al. gave a hamiltonian formulation of their 
network model\cite{Kagalovsky}.  The result is the $4\times4$ matrix
hamiltonian:
\beq
\label{h1}
H = ( \tau_x p_x + \tau_z p_y ) \otimes 1 + 1 \otimes \alvec\cdot
\sigvec
\eeq
where $\vec{\tau}, \sigvec$ are two copies of the Pauli matrices, 
$p_{x,y} = -i \d_{x,y}$, and $\alvec$ is a random spin potential. 
This hamiltonian has the defining properties to belong
the class C of the classification introduced in ref.\cite{AlZirn}. 
Let us perform a unitary transformation $H \to (U^\dagger \otimes 1) H
(U \otimes 1)$, where $U$ corresponds to a rotation about the $x$ axis 
for the Pauli matrices:  
$U^\dagger \tau_z U = \tau_y$, $U^\dagger \tau_y U = - \tau_z $, 
$U^\dagger \tau_x U = \tau_x$.   After including $su(2)$ gauge
potentials to $\vec{p}$,  one has the following $2\times 2$ block 
structure:
\begin{equation}
\label{2.1}
H = \left( \matrix{\alvec\cdot\sigvec & -i \d_\zbar + A_\zbar\cr
-i \d_z + A_z & \alvec\cdot \sigvec \cr} \right) 
\end{equation}
where $\d_z = \d_x - i \d_y $, $\d_\zbar = \d_x + i\d_y $, 
$A_z =  A_x - i A_y $,   $A_\zbar = A_x + i A_y $.
In the above hamiltonian, $\alvec(x,y)$ is a real, random spin
potential, and $A_\mu = \sum_a A_\mu^a (x,y) \sigma^a $ are 
random $su(2)$ gauge potentials, with $A^a_{x,y}$ real. 
In the sequel we will take them to have the following gaussian
distributions

\def\dx{ \frac{d^2 x}{2\pi} }

\begin{eqnarray}
\label{2.2}
P(\alvec ) &=& \exp \( -\inv{g_\al} \int  
\dx ~  \alvec(x)  \cdot \alvec(x) \) 
\\ \nonumber 
P(A_\mu) &=& \exp \( - \frac{4}{g_A} \int \dx ~ A_z^a (x) A_\zbar^a (x) \) 
\end{eqnarray}

As we will show in the next section, renormalization of the effective action
obtained upon disorder averaging leads to an additional interaction which can
be viewed as arising from a random mass $m(x,y)$.  The complete hamiltonian
which leads to a renormalizable effective action is then 

\begin{equation}
\label{2.3}
H = \left( \matrix{\alvec\cdot\sigvec + m & -i \d_\zbar + A_\zbar\cr
-i \d_z + A_z & \alvec\cdot \sigvec -m \cr} \right) 
\end{equation}
We will take $m(x)$ to have the gaussian distribution 
\beq
\label{2.4} 
P(m) = \exp \( -\inv{g_m} \int \dx  ~ m(x)^2  \) 
\eeq
Note that since $\d_z^\dagger = -\d_\zbar$, $A_z^\dagger = A_\zbar$,
the hamiltonian is hermitian if $m , \alvec, A_{x,y}$ are real.  
In this situation the couplings $g_{\al} , g_m , g_A $ are the  variances
of normalizable gaussian distributions if they are real and positive.  
Negative couplings $g_{\al, m, A}$ can be interpreted as corresponding 
to imaginary random potentials;  hermitian hamiltonians can then be constructed
by doubling the number of degrees of freedom, as in \cite{GLL}.  

\def\psib{\bar{\psi}}

The single particle Green functions are defined by the functional integral
$Z^{-1}\,\int D\Psi^* D\Psi ~ \exp (-S)$ with $Z$ the partition function and
\beq
\label{2.5} 
S = \int \dx ~ \Psi^{\star} (x) i \( H - \CE \) \Psi (x) 
\eeq
where $\CE = E + i\vep$.  For $\vep = 0^+$, this defines the retarded 
Green function
\beq
\label{2.6}
G_R (x,x';E) = \lim_{\vep \to 0^+} \langle x | \inv{H-(E+i\vep)} |x' \rangle 
= \lim_{\vep \to 0^+} i \langle \Psi (x) \Psi^{\star} (x') \rangle 
\eeq
Letting 
\beq
\Psi = \left( \matrix{ \psib_+ \cr \psi_+ \cr} \right), 
 ~  \Psi^{\star} = (\psi_- , \psib_- ) 
\eeq
where $\psi_+$ is a 2-component fermion $\psi_+^i, i=1,2$, 
and similarly for $\psib_\pm$, one finds 
\begin{eqnarray}
\nonumber
S &=& \int \dx \BL \psi_- 
( \d_\zbar + i A_\zbar ) \psi_+  + \psib_- (\d_z + i A_z ) \psib_+ 
+ i\alvec \cdot (\psi_- \sigvec \psib_+ + \psib_- \sigvec \psi_+ ) \non
\\  
&~& ~~~~~~~~~~~~~~  + im (\psi_- \psib_+ - \psib_- \psi_+ ) 
-i \CE \Phi_E   \BR 
\label{2.8}
\end{eqnarray} 
where 
$$\Phi_E = \psi_- \psib_+ + \psib_- \psi_+ $$  
(We have suppressed the $su(2)$ indices, i.e. 
$\psi_- \psib_+ = \sum_{i} \psi_-^i \psib_+^i $, etc.) 

In \cite{Kagalovsky} the perturbation $H_\ep = \ep \,  \tau_y \otimes 1$ was
added and critical exponents for $\ep$ were measured numerically. 
Performing the unitary transformation $U$ defined above, this results
in the following perturbation of the action: 
\beq
\label{sep}
S_\ep = -i \ep\, \int \dx  \Phi_\ep , ~~~~~~~~ \Phi_\ep = 
\psi_- \psib_+ - \psib_- \psi_+  
\eeq
Non-zero $\ep$ corresponds to a non-zero average random mass $m$.

Let us attempt to compare this with the model considered by 
Senthil et. al.\cite{Senthil}.  
There, one had a one-dimensional lattice in the $x$-direction with sites labeled
by $j$, and a continuous $y$ direction.  
At each site there are fermionic degrees of freedom $\chi_j (y)$, where $j$ even
corresponds to left-movers and $j$ odd to right-movers.  The hamiltonian is 
\begin{eqnarray}
\nonumber
H &=& \sum_j \int dy \BL 
(-)^j \chi^\dagger_j (y) ( -i\d_y + \vec{\eta}_j (y) \cdot \sigvec ) \chi_j (y) 
-i t^0_j (y) (\chi^\dagger_{j+1} \chi_j - \chi_j^\dagger \chi_{j+1} ) 
\\ \label{2.9}  
&~& ~~~~~~~~~~~~~~~~~~~~~
+\vec{t}_j (y) \cdot( \chi^\dagger_{j+1} \sigvec \chi_j + \chi_j^\dagger \sigvec\chi_{j+1}
) \BR 
\end{eqnarray}
Taking a continuum limit,  $\chi_j (y) \to \psi_+ (x,y) $ for $j$ even and $\chi_j (y) \to
\psib_+ (x,y) $ for $j$ odd,  $\vec{\eta}_j (y) \to \vec{\eta}_y (x,y)$, and
$t_j (y) \to t (x,y)$ , one finds
\begin{eqnarray}
\nonumber
i H &=& \int dx dy \BL \psi_- (\d_y +i \vec{\eta}_y \cdot \sigvec ) \psi_+ 
- \psib_- (\d_y + i \vec{\eta}_y \cdot \sigvec ) \psib_+ 
+ t_0 \, ( \psi_- \psib_+ - \psib_- \psi_+ ) 
\\ 
&~& ~~~~~~~~~~~~~~~~~~
+ i \vec{t} \cdot (\psi_- \sigvec \psib_+ + \psib_- \sigvec \psi_+ ) \BR 
\label{2.10}
\end{eqnarray}
Rotational invariance of the kinetic terms can be restored by adding 
$\d_x + i \vec{\eta}_x \cdot \sigvec$ to the derivatives.  Performing a rotation
to euclidean space $y \to -iy $, one then obtains the action (\ref{2.8}) 
with $\alvec = \vec{t}$, $A_z = (\vec{\eta}_x - i \vec{\eta}_y)\cdot\sigvec$ 
and $t_0 = im$.  

In \cite{Senthil}, $\vec{t}$ and $t_0$ were taken as real gaussian 
distributed but with the {\it same} variance.  For the model defined by 
(\ref{2.8}), this corresponds to imaginary $m$.  Letting $m \to -i m$, one sees
that for the effective theory obtained after disorder averaging, imaginary $m$
corresponds to negative $g_m$.  Since the variances are the same, this corresponds
to $g_\alpha = - g_m$.\footnote{This 
appears to be inconsistent with hermiticity,  since the hamiltonian (\ref{2.3}) 
is not hermitian for imaginary $m$, suggesting the two models cannot be
naively compared in this way.}      
In summary, though we have not described an exact
mapping between the  network model in \cite{Kagalovsky} and the super spin chain
in
\cite{Senthil,Read},    
it  appears that they should be related on the line 
$g_\alpha + g_m = 0$.  In fact, one obtains directly
the Heisenberg type superspin chain upon disorder averaging 
only along this line\cite{Senthil}.     
We will provide 
further support of this statement based on symmetry in the sequel. 
As we will show, this line
has some rather special properties which allow the model to be solved.

\section{Effective Action and Renormalization Group Analysis} 

\subsection{Supersymmetric Disorder Averaging}

\def\betab{\bar{\beta}}

Since we are dealing with a free field theory, the supersymmetric method for
disorder averaging can be used.  Conformal field theory techniques in
 conjunction with this method was used for other models in \cite{carg}\cite{mudry}. 
We augment the theory with bosonic ghosts
$\beta^i_\pm, \betab_\pm^i $, $i=1,2$, so that the inverse of the fermionic
partition function $Z (\alvec, m, A) = \int D\psi e^{-S(\psi)} $ 
is represented as a bosonic functional integral: 
\beq
\label{3.1}
Z(\alvec, m , A)^{-1}  = \int D\beta ~ e^{-S(\psi \to \beta)} 
\eeq
One can then perform the gaussian integrals over the random potentials.  The 
result is the effective action:
\beq
\label{3.2} 
S_{eff} = S_{cft} + \int \dx \( g_\alpha \, \CO_\al  + g_m \, \CO_m + g_A \, \CO_A
\) 
\eeq
The conformal field theory $S_{cft}$ has Virasoro central charge $c=0$ and
the action:
\beq
\label{3.3} 
S_{cft} = \int \dx \( \psi_- \d_\zbar \psi_+ + \psib_- \d_z \psib_+ 
+ \beta_- \d_\zbar \beta_+ + \betab_- \d_z \betab_+ \) 
\eeq
The first-order action for the bosonic ghosts of conformal dimension $\pm 1/2$ 
can be treated as in \cite{FMS}.   The  operators  which perturb away from
the conformal field theory are: 
\begin{eqnarray}
\nonumber
\CO_\al &=& \inv{4} \( \psi_- \sigvec \psib_+ + \psib_- \sigvec \psi_+ 
+ \beta_- \sigvec \betab_+ + \betab_- \sigvec \beta_+ \)^2 
\\ 
\label{3.4}
\CO_m &=& \inv{4} \( \psi_- \psib_+ - \psib_- \psi_+ + \beta_- \betab_+ - \betab_-
\beta_+ \)^2 
\\  
\nonumber
\CO_A &=& \inv{4} \(\psi_- \sigvec \psi_+ + \beta_- \sigvec \beta_+ \) \cdot
\(\psib_- \sigvec \psib_+ + \betab_- \sigvec \betab_+ \)  
\end{eqnarray}

\subsection{$\beta$eta Functions} 

The one-loop renormalization group $\beta$eta functions can be deduced from the 
operator product expansions of the marginal perturbing operators $\CO_i$\cite{Zamo}.
Namely, if 
\beq
\label{3.5} 
\CO_i (z, \zbar) \CO_j (0)  \sim \inv{z\zbar} C_{ij}^k ~ \CO_k (0) + ....
\eeq
then to lowest order the $\beta$eta function  $\beta_k = dg_k / d \log l $, where
$l$ is a length scale,  is given by 
\beq
\label{3.6} 
\beta_k = - \sum_{i,j} C^k_{ij}  ~ g_i g_j 
\eeq

We normalize the Pauli matrices as follows:  
$\sigvec \cdot \sigvec = (\sigma^3)^2 + \sigma^+ \sigma^- + \sigma^- \sigma^+ $, 
with 
\beq
\label{3.8} 
\sigma^3 = \left( \matrix{1&0\cr 0&-1 \cr } \right), ~~~~~ 
\sigma^+ = \sqrt{2} \left( \matrix{0&1\cr 0&0 \cr } \right), ~~~~~ 
\sigma^- = \sqrt{2} \left( \matrix{0&0\cr 1&0 \cr } \right) 
\eeq
With this normalization, one has  $Tr \sigma^a \sigma^b = 2 \delta^{ab}$, 
and $f^{abc} f^{abd} = -8 \delta^{cd}$, where  
$[\sigma^a , \sigma^b] = f^{abc} \sigma^c$.  
One also needs the operator products:
\begin{eqnarray}
&~&\psi_+ (z) \psi_- (0) \sim \psi_- (z) \psi_+ (0) \sim  1/z 
\\ \nonumber
&~&\beta_+ (z) \beta_- (0) \sim - \beta_- (z) \beta_+ (0) \sim 1/z 
\label{3.7} 
\end{eqnarray}
and similarly for the right-movers $\psib, \betab$ with $z$ replaced by 
$\zbar$. 
One then finds 
\begin{eqnarray}
\CO_A (z,\zbar) \CO_A (0) &\sim& \frac{-2}{z\zbar} \CO_A  , ~~~~~
\CO_m (z,\zbar) \CO_m (0) \sim \frac{2}{z\zbar} \CO_m 
\\ \nonumber 
\CO_\al (z,\zbar) \CO_\al (0) &\sim& \inv{z\zbar} \( 2 \CO_\al  - 8 \CO_A  \) 
\\ \nonumber
\CO_A (z,\zbar) \CO_\al (0) &\sim& \inv{z\zbar} \( -\inv{2} \CO_\al  - \frac{3}{2}
\CO_m  \) 
\\ \nonumber
\CO_A (z,\zbar) \CO_m (0) &\sim& \inv{z\zbar} \( -\frac{3}{2}  \CO_m  - \frac{1}{2}
\CO_\alpha   \) 
\\ \nonumber
\CO_\al (z,\zbar) \CO_m (0) &\sim& \inv{z\zbar} \( 3  \CO_m  - 
\CO_\alpha    - 4 \CO_A  \) 
\label{3.9}
\end{eqnarray}
The resulting $\beta$eta functions are
\begin{eqnarray}
\nonumber
\beta_\al &=&  - 2 g_\al^2 +  2 g_\al g_m +  g_A (g_\al + g_m)  \\
\label{3.10}
\beta_m&=&   - 2g_m^2  - 6 g_\al g_m  + 3 g_A (g_\al + g_m ) \\
\nonumber
\beta_A  &=& 2g_A^2 + 8 g_\al ( g_\al + g_m )  
\end{eqnarray}

If one starts with a model with only the random spin potential 
$\alvec$, i.e. only $g_\al \neq 0$, then $g_A$ and $g_m$ are generated under
renormalization.  Note that the $\beta$eta functions simplify dramatically 
when $g_\al + g_m = 0$, which is the subject of the next section.  
We will return to a general analysis of the possible universality classes
contained in these $\beta$eta functions in section V.

\section{Spin-Charge Separation } 

\def\osp{osp(2|2)}
\def\Sh{\hat{S}}
\def\Jb{\bar{J}} 
\def\Sb{\bar{S}}
\def\Hb{\bar{H}}

\subsection{Symmetries of the effective action} 

The free conformal action $S_{cft}$ has a maximal $osp(4|4)_1$ 
current algebra symmetry. 
The conformal currents $J^A$  generating this symmetry correspond to all
possible bilinears in fermions and ghosts. Suppressing the possible index
structure,  $J^A = \{ \psi \psi, \psi \beta,\beta\beta \}$.  
The perturbing operators  are bilinear in these currents:
$\CO_\al=d^{(\al)}_{AB}\, J^A\bar J^B$ for some tensor $d^{(\al)}_{AB}$
and similarly for $\CO_m$ and $\CO_A$.

For arbitrary $g_\alpha, g_m , g_A$,  the $osp(4|4)$ symmetry of
$S_{cft}$ is broken but an $\osp$ symmetry is preserved.  
To describe the symmetry, let us define the left-moving currents:
\begin{eqnarray}
\nonumber
J &=& \sum_i  \psi_+^i \psi_-^i  ,  ~~~~~~~~~~
J_\pm  = \sum_{i,j} \ep_{ij} \psi^i_\pm \psi^j_\pm
\\ \label{4.2} 
H &=& \sum_i \beta_+^i \beta_-^i
, ~~~~~~~~~~
S_\pm =  \sum_i \psi_\pm^i \beta_\mp^i  
\\  
\hat{S}_\pm  &=& \sum_{i,j}  \ep_{ij} \psi^i_\pm \beta^j_\pm   \nonumber
\end{eqnarray}
where $\ep_{ij} = -\ep_{ji}$, $\ep_{12} = 1$, and similarly for the right-movers
$\bar{J} = \psib_+ \psib_-,  \bar{H} = \betab_+ \betab_-, .....$.  
These currents generate an $\osp_k$ current algebra at level $k=-2$.  The non-zero
operator products are 
\begin{eqnarray}
\nonumber
J(z) J(0) &\sim& - \frac{k}{z^2} ,  ~~~~~~~~~~~~~~~H(z) H(0) \sim \frac{k}{z^2} 
\\ \nonumber
J(z) J_\pm (0) &\sim&  \pm \frac{2}{z} ~ J_\pm   
, ~~~~~~~~~~
J_+(z) J_- (0) \sim \frac{2k}{z^2} - \frac{4}{z} J 
\\ \nonumber
J(z) S_\pm (0) &\sim&  \pm \inv{z} S_\pm   ,  ~~~~~~~~~~ J(z) \Sh_\pm (0) \sim
\pm \inv{z} \Sh_\pm  
\\ \nonumber
H(z) S_\pm (0) &\sim&  \pm \inv{z} S_\pm  , ~~~~~~~~~~
H(z) \Sh_\pm (0) \sim \mp \inv{z} \Sh_\pm 
\\ \nonumber
J_\pm (z) S_\mp (0) &\sim& \frac{2}{z} \Sh_\pm , ~~~~~~~~~~~~~~~
J_\pm (z) \Sh_\mp (0) \sim - \frac{2}{z} S_\pm  
\\ 
\label{4.3} 
S_\pm (z) \Sh_\pm (0) &\sim& \pm \inv{z} J_\pm 
\\ \nonumber
S_+ (z) S_- (0) &\sim&  \frac{k}{z^2} + \inv{z} (H-J) 
\\ \nonumber
\Sh_+ (z) \Sh_- (0) &\sim& - \frac{k}{z^2} + \inv{z} (H+ J) 
\end{eqnarray}
The currents $J$ and $J_\pm$ generate a charged $su(2)$ subalgebra under
which the fermions $\psi^i_\pm$ transform as doublets.

The $\osp$ symmetry is present for arbitrary potentials 
before disorder averaging, as we now describe.
The fermionic generators define nilpotent transformations of
the bosonic and fermionic fields. The action (\ref{2.8}) is then
an exact variation with respect to these transformations, making
then clear the fact that this $\osp$ symmetry 
is preserved.  
For example the generator $S_+$ induces the transformation
$\de \psi_+^i=0$, $\de \psi^i_-=\beta_-^i$,  
$\de \beta_+^i=-\psi^i_+$, $\de \beta_-^i=0$.  The symmetry acts 
left-right diagonally, so  
for the right movers $\delta\psib^i_+ = 0, 
\delta \psib_-^i = \bar{\beta}_-^i$, $\delta \bar{\beta}^i_+ = - 
\psib_+^i , \delta \bar{\beta}_-^i = 0$.  The action 
(\ref{2.8}) may be written as
$$
S = S_{cft} + \de\, \int \dx \Th
$$
where
\beq
\label{theta}
\Th = i m ( \psi_- \betab_+ - \psib_- \beta_+ ) 
+ i\alvec \cdot (\psi_- \sigvec \betab_+ + \psib_- \sigvec \beta_+ ) 
+ i \vec{A} \cdot (\psi_- \sigvec \beta_+ + \psib_- \sigvec \betab_+ ) 
\eeq
Hence $\delta S = 0$ since $\delta S_{cft} = 0$ and $\delta^2 = 0$.  
This holds similarly, but with another operator $\hat \Th$,
for $\hat S_-$ which generates the transformations
$\hat \de \psi_+^i=\ep^{ij}\beta_-^j$, $\hat \de \psi_-^i=0$
and $\hat \de \beta_+^i=\ep^{ij}\psi_-^j$, $\hat \de \beta_-^i=0$.
This $\osp$ symmetry would not be preserved if
we had chosen the random gauge potential in $u(2)$ or if
had added extra scalar potential randomness.

This implies that after disorder averaging  the perturbations by $\CO_\al$,
$\CO_m$ and $\CO_A$ in the effective theory 
preserve a global, left-right diagonal,  $\osp$ symmetry. 
The current conservation law takes the left-right diagonal form 
\beq
\label{4.6} 
\d_\zbar J^a + \d_z \Jb^a = 0 
\eeq
for $J^a, \Jb^a $  any of the eight  $\osp_{-2}$ currents.

Let us describe more explicitly how the $\osp$ symmetry is manifested in
the effective theory.  In the sequel we will set $g_\al = -g_m = g$ and
the effective action will contain the operator $\CO_g \equiv \CO_\al - \CO_m$. 
Many terms cancel in the combination $\CO_\al - \CO_m$
and the result can be written as an $osp(2|2)$ current-current perturbation.
This simplification is analogous to the $g_V + g_M = 0$ line for the
random $U(1)$ fermions which has $gl(1|1)$ symmetry,  studied in \cite{GLL}.
By repeated use of the identity 
\beq
\label{4.4} 
\sigma^a_{ij} \sigma^a_{nm} = 2 \delta_{im} \delta_{jn} - \delta_{ij} \delta_{nm} 
\eeq
one can express $\CO_g$ in terms of 
the above $\osp$ currents: 
\debut
\label{4.5}
\CO_g &\equiv &\CO_\al - \CO_m \\
&=&  - J\Jb + H\Hb + \inv{2} (J_- \Jb_+ + J_+ \Jb_- ) 
+ S_- \Sb_+ - S_+ \Sb_- - \Sh_- \bar{\Sh}_+  + \Sh_+ \bar{\Sh}_- \non
\fin
The operator $\CO_g$ has the structure of the quadratic Casimir for $\osp$.  
Namely,  $\CO_g = \sum_{a,b} C_{ab} J^a \Jb^b $, where $J^a$ are $\osp$ currents
and $C_{ab}$ corresponds to the quadratic Casimir.  Thus $\CO_g$ is $\osp$ invariant.

The operator $\CO_\al+\CO_m $ cannot be written only in terms
of the $\osp$ currents; one needs some of the    $osp(4|4)$ currents.
Let
\debut
B^{ij}_\pm = \beta_\pm^i\beta_\pm^j
\quad &;&\quad 
\vec L_f =\psi_- \vec\sig\psi_+ ~~ ; ~~ \vec L_b = \beta_-\vec\sig\beta_+ \non\\
U^{ij}_\pm =\half( \beta_\pm^i\psi_\pm^j + \beta_\pm^j\psi^i_\pm)
\quad &;&\quad 
\vec V_- = \psi_-\vec\sig\beta_+ ~~ ; ~~ \vec V_+ = \beta_-\vec\sig\psi_+
\fin
Then, $$\CO_\al+\CO_m = \half \CO_g  + \tilde{\CO} $$ with
\debut
\tilde{\CO}  &=&\half\( B_-^{ij}\bar B_+^{ij} + B_+^{ij}\bar B_-^{ij}\)
+\half\(\vec L_f\cdot {\vec {\bar L}}_f - 
\vec L_b\cdot {\vec {\bar L}}_b \) \non\\
&~&+U_-^{ij}\bar U_+^{ij} - U_+^{ij}\bar U_-^{ij} 
+ \half\( \vec V_+\cdot {\vec {\bar V}}_- - \vec V_-\cdot{\vec {\bar V}}_+ \)
\fin
We have already seen that $\CO_g$ is $\osp$ invariant.  One can check 
explicitly that $\tilde{\CO}$ is also $\osp$ invariant.  

Let us turn now to the perturbation $\CO_A$.  This operator as defined is 
a left-right current-current perturbation:
\beq
\label{4.7} 
\CO_A = \sum_{a=1}^3  L^a \bar{L}^a 
\eeq
where 
\beq
\label{4.8} 
L^a = L_f^a + L_b^a = \psi_- \sigma^a \psi_+ + \beta_- \sigma^a \beta_+ 
\eeq
The central extension (level) cancels between the fermions and ghosts and the
result is that $L^a$ generate an $su(2)_k$ current algebra at level $k=0$: 
\beq
\label{4.9}
L^a (z) L^b (0) \sim \inv{z}  f^{abc} L^c (0) +  {\rm reg.} 
\eeq
Again, since $\CO_A$ takes the form of the Casimir of $su(2)$, $\CO_A$ preserves
a global $su(2)$ symmetry.  Furthermore it is easy to check that this
global $su(2)$ commutes with the $\osp$ symmetry:
$$ \[ L^a , J^b \] = 0$$
where $J^a$ are the $\osp$ currents defined in eq. (\ref{4.2}).   

In summary, the model has a global $\osp \otimes su(2)$ symmetry  
for general $g_\al , g_m , g_A$.

\subsection{Quasi-spin-charge separation in the conformal field theory} 

Since $\CO_g$ only involves the $\osp_{-2}$ currents and $\CO_A$ the 
$su(2)_0$ currents, it is important to understand how the conformal field
theory decomposes in terms of these current algebras.  It is straightforward
to check that the $su(2)_0$ and $\osp_{-2}$ current algebras commute, so that
the conformal field theory  defined by $S_{cft}$ in eq. (\ref{3.3}) has 
an $\osp_{-2} \otimes su(2)_0$ symmetry.  

We now show the much stronger result
that the full stress tensor, which is the Sugawara 
stress tensor for the   $osp(4|4)_1$ supercurrent algebra, 
separates into two commuting pieces:
\beq
\label{4.11}
T_{osp(4|4)_1}  = T_{\osp_{-2}} + T_{su(2)_0} 
\eeq
All $T$'s in the above equation are Sugawara stress tensors
and  have Virasoro central charge equal to zero.  
The above equation is proven in the appendix.  

Since $\osp_{-2}$ contains charged currents whereas $su(2)_0$ does not, the
equation (\ref{4.11})  implies a kind of spin-charge separation and we will
use this terminology in the sequel.  

Based on the separation (\ref{4.11}) one would expect that the
Hilbert space of $osp(4|4)_1$ factorizes.  However by trying to
perform explicitly this factorization in some simple cases, we
found that it is not possible: 
$$
\CH_{osp(4|4)_1}  \neq  \CH_{osp(2|2)_{-2}}\ \otimes \ \CH_{ su(2)_0 }
$$
This non-factorization is intimitely related to the way the
fields that appear in $\tilde{\CO}$ transform under $osp(2|2)$.
This is described in Figure 2.
In particular, the $su(2)_0$ currents $L^a$ turn out to
be susy exact, e.g.:
\beq
\label{factor}
S_+ (z) V^a_- (0)  \sim \inv{z} L^a 
\eeq

Let us argue for the non-factorazibility of the Hilbert space by
contradiction. 
The $24$ fields of $\tilde{\CO}$ form three copies of an eight
dimensional reducible but indecomposable representation
of $osp(2|2)$, that we shall denote $[\tilde 8]$.
Consider the state $\ket{L^a_A}\equiv (L^a_{f,-1}-L^a_{b,-1})\ket{0}$,
where $L^a_{-1}  = \oint dz  ~ L^a (z) / 2i \pi  z$.
If the factorization of $\CH_{osp(4|4)_1}$ holds, then
$$
\ket{L^a_A}=\sum_i \ket{osp}_i\otimes \ket{su}_i 
\in [\tilde 8]\otimes [3]
$$
where $[3]$ is the adjoint representation of $su(2)$. 
Let us now act on $\ket{L^a_A}$ with the $su(2)_0$ current. 
As a state in $\CH_{osp(4|4)_1}$ one finds that
$L^b_{-1}\ket{L^a_A}=\delta^{ab}\,\ket{0_{\rm Fock}}$ 
with $\ket{0_{\rm Fock}}$ the vacuum of the Fock space.
As a state in $\CH_{osp(2|2)_{-2}}\ \otimes \ \CH_{ su(2)_0 }$, one
would find 
$$
L^b_{-1}\ket{L^a_A}=\sum \ket{osp}_i\otimes L^b_{-1}\ket{su}_i 
\in [\tilde 8]\otimes [1]
$$
where $[1]$ is the $su(2)$ singlet. 
Thus if the factorization  holds, one would deduce that
$$
\ket{0_{\rm Fock}} \in [\tilde 8]\otimes [1]
$$
Since $\ket{0_{\rm Fock}}$ is annihilated by all $osp(2|2)$ generators,
this would mean that $\ket{0_{\rm Fock}}$ is in the image of susy
generators, i.e. there would exist a state $\ket{\Om}$ such that for
example 
$\ket{0_{\rm Fock}}=S_-\ket{\Om}$.  This is clearly a contradiction.
A similar argument shows that eq. (\ref{factor}) leads to a contradiction
in the factorization of $V_-^a$.  

Another argument for the non-factorization is based on the fact that
the current algebras $osp(2|2)_{-2}$ and $su(2)_0$ lead to logarithmic
correlation functions\cite{Serban}\cite{tsvel2}, 
whereas the $osp(4|4)_1$ is a free theory with
no logarithms. 
It appears impossible for these logarithms to cancel
in the product of two logarithmic correlation functions.   
The logarithmic nature of these theories can
in fact be traced to transformations such as 
(\ref{factor})\cite{Serban}.

We now describe a few of the results we need concerning these current algebra
conformal field theories.  Representations of $\osp$ are characterized by the quantum
numbers of the $su(2) \otimes u(1)$ subalgebra generated by $(J, J_\pm )$ 
and $H$.  Highest weights  can be labeled $(j,b)$ where $j=0,1/2, 1, ..$ is the spin
of the charge-$su(2)$ and $b=H/2$.  These representations have dimension $8j$.  
The conformal scaling dimension (left or right-moving) 
of the corresponding primary fields, as
determined from the Sugarawa construction,  is \cite{Bowcock,Serban}
\beq
\label{4.13}
\Delta^{osp(2|2)}_{(j,b)} = \frac{2(j^2 - b^2)}{2-k}  
\eeq
On the other hand, the primary fields of the $su(2)_k$ current algebra are
characterized by the   spin $j$ of the spin-$su(2)$ only and have conformal dimension\cite{KZ} 
\beq
\label{4.14}
\Delta^{su(2)}_j  = \frac{j(j+1)}{k+2} 
\eeq

A simple check of eq. (\ref{4.11}) is based on the dimensions of the eight
original fermion and ghost fields $\psi_\pm^i , \beta_\pm^i$.  In the original
conformal field theory these fields have conformal dimension $\Delta = 1/2$.  
Under the $\osp$  they transform according to the two four-dimensional
representations  $(\psi_+^1 , \psi_-^2 , \beta_-^2 , \beta_+^1 )$ 
and $(\psi_-^1 , \psi_+^2 , \beta_+^2 , \beta_-^1 )$ corresponding to 
$(j=1/2, b=0)$ and its conjugate.  At level $k=-2$, these have dimension 
$\Delta^{\osp} = 1/8$.  
The  same fields transform as spin $j=1/2$ doublets according to the $su(2)_0$,
and have dimension $\Delta^{su(2)} = 3/8$ at level zero.  The exact decomposition
(\ref{4.11}) implies that these dimensions must add up properly:  $1/2 = 1/8 + 3/8$.

\subsection{Infrared  fixed points and critical exponents}

We now study the model when $g_\al + g_m = 0$.  
Setting $g_\al = -g_m = g$, the effective action  contains the 
current-current operators $\CO_g$ and $\CO_A$.  The operator
$\tilde{\CO}$ which couples the $\osp_{-2}$ and $su(2)_0$ current algebras
is not present.  
The model in \cite{Senthil} was mapped onto a super-spin chain with 
$osp(2|2)$ symmetry and Heisenberg-type hamiltonian.
This further supports the  identification of
this super spin chain   with our   model on the line $g_\al + g_m = 0$, 
since,  at least for ordinary bosonic algebras, Heisenberg hamiltonians correspond
to symmetry preserving current-current perturbations built on the quadratic casimir
in the continuum limit.  
(As Lie superalgebras, $\osp$ and $sl(2|1)$ are identical.)

The effective action
now contains interactions that do not couple the spin and charge currents:
\beq
\label{4.15}
S_{eff} = S_{\rm cft}  
+ \int \dx ~\({\, g \CO_g +  g_A \CO_A \,}\)
\eeq 
This decoupling is of course consistent with the more general $\beta$eta functions
(\ref{3.10}).  Setting $g_\al = - g_m = g$ one finds 
\beq
\label{4.17}
\beta_g = -4 g^2 , ~~~~~ \beta_{g_A} = 2 g_A^2 
\eeq

As explained in section II, the model of Senthil et. al. corresponds to $g$ and 
$g_A$ positive.  From the above $\beta$eta functions one sees that $g_A$ is
marginally relevant whereas $g$ is marginally irrelevant.   In order to 
deduce something exact concerning the infrared (IR) fixed point theory from
the one-loop $\beta$eta functions,  one
needs to make a hypothesis concerning the role of the higher loop corrections.  
Consider for comparison  the $su(N)$ Gross-Neveu models, which are $su(N)$ 
current-current perturbations,   with $\beta$eta functions
as in eq. (\ref{4.17}), i.e. $\beta_g = g^2$  in a certain convention.  
When $g>0$ the perturbation is marginally relevant, i.e. $g$ grows at large
distances.  Higher loop corrections do not modify this, i.e. the flow to
the IR does not stop at some finite value of $g$ corresponding to a non-trivial
fixed point.  Rather,  $g$ eventually
flows to infinity.  The theory is thus  massive, and in the infrared all these massive modes
disappear leaving no massless degrees of freedom,  i.e. the infrared theory is
an empty theory.  When $g<0$, the perturbation is marginally irrelevant,
i.e. $g$ flows back to zero and the unperturbed conformal current algebra is
recovered in the IR.  We will make the hypothesis that this is the only possible
behavior for general current-current perturbations, i.e. the only possible fixed
points are $g=0$ or $\infty$.  

For our model we then have the following picture.  In the IR, $g_A$ flows to 
infinity and the spin degrees of freedom are massive and decouple.  
The coupling  $g$ on the other hand   is marginally
irrelevant.  Therefore the IR fixed point is the coset 
$osp(4|4)_1 /su(2)_0$,
and the theory arrives in the IR via the operator $\CO_g$.   
Due to the non-factorizability of the Hilbert space, this coset conformal
field theory is not precisely the $osp(2|2)_{-2}$ current algebra, 
even though it possesses this current algebra as a symmetry and
the conformal dimensions of the coset theory are the same as 
the $osp(2|2)$ current algebra because of (\ref{4.11}). 

This identification of the IR fixed point allows the computation of certain 
critical exponents.  The density of states $\rho (E)$ is 
\beq
\label{4.18}
\rho(E) = \inv{V} {\rm Tr}\, \delta (H-E) = \inv{\pi V} \lim_{\vep \to 0^+} 
{\rm Im} \, Tr \inv{H-E - i \vep} 
\eeq
where $V$ is the two-dimensional volume.  This implies that the disorder
averaged density of states is proportional to the one-point 
correlation function:
\beq
\label{4.19}
\bar{\rho(E)} \propto\, {\rm Re}\, \langle \Phi_E \rangle
\eeq
Let $\Gamma_E$ equal the scaling dimension of $\Phi_E$.  Then, since the action 
(\ref{2.8}) is
dimensionless,  viewing $E$  as a  coupling,
${\rm dim}(E) = 2-\Gamma_E$.  Since $E$ is the only dimensionful
coupling in the theory one deduces  
\beq
\label{4.20}
\bar{\rho(E)} \propto  E^{\Gamma_E /(2-\Gamma_E )} 
\quad {\rm as}\quad E\to 0
\eeq
One can also define a correlation length $\xi_E$, 
\beq
\label{4.21}
\xi_E \propto E^{-\nu_E}, ~~~~~~~~~~~~ \nu_E = 1/(2-\Gamma_E )
\eeq
(In \cite{Senthil}, $\nu_E$ was referred to as $\nu_B$.)

For our theory, $\Phi_E = \psi_+ \psib_- + ...$ and $\Gamma_E$ is the scaling
dimension in the IR, which follows from the $osp(2|2)_{-2}$ conformal dimension of
$\psi_\pm$, which as explained above is $\Delta^{\osp}_{(1/2, 0)} = 1/8$.  
Thus $\Gamma_E = 1/4$, and:
\beq
\label{exp1}
\bar{\rho(E)} \propto E^{1/7}, ~~~~~~~~~ 
\nu_E = 4/7  
\eeq
Numerical simulations of the $\osp$ invariant spin-chain  
agree very well with $\nu_E =4/7$ \cite{Senthil}.
Since the system flows toward the infrared fixed point along
a marginal direction, the scaling (\ref{exp1}) is
up to computable logarithmic corrections.

In the above scenario, the $su(2)_0$ spin degrees of freedom are becoming
localized, hence the terminology ``spin quantum hall effect''.  
The above model can also exhibit charge localization by changing the signs
of the couplings.  Namely, if $g$ and $g_A$ are negative, then $g$ is marginally
relevant, i.e. it flows to $-\infty$ in the IR and $g_A$ is marginally
irrelevant.  Following the same reasoning as above, in this case 
the charge degrees of freedom  decouple in the flow.
 The IR fixed point is now the coset $osp(4|4)_1/osp(2|2)_{-2}$.  
 Here, $\Gamma_E = 3/4$, and: 
\beq
\label{exp2}
\bar{\rho(E)} \propto E^{3/5} , ~~~~~~~~~~~~  \xi_E \propto E^{-4/5}   
~~~{\rm for ~ charge ~ delocalization} 
\eeq

Finally, if $g<0, g_A > 0$, then $g,g_A$ flow to $-\infty$ and $\infty$ 
respectively, and both sectors are massive and decoupled in the IR.
This implies $\Gamma_E = 0$, and a constant density of states at $E=0$.

\subsection{Path Integral Factorization}
Let us now present a path integral derivation of the spin-charge
separation which has the advantage of being non perturbative and thus valid
all along the RG trajectory. 
One can view this formulation as a way of defining the coset
$osp(4|4)_1/su(2)_0$. 
It consists in decoupling the random gauge field  $A$
using chiral gauge transformations. 
The spin-charge separation will then be a consequence of the fact 
that the $osp(2|2)$ currents are invariant under the chiral
gauge transformations. This decoupling is similar to the solution of the
random gauge potential described in \cite{carg}.

Let $S$  be the fermionic action (\ref{2.8})
and $Z(A,\al,m)$ its partition function
 $Z(A,\al,m)= \int D\Psi\ e^{-S(\Psi)}$.
At a fixed  realization of disorder,  the gauge potential $A$ may be gauged
away by a chiral gauge transformation by parameterizing $A$ as:
\debut
i \bar {A} = G^{-1}\d_{\bar z} G \quad;\quad i A = G^{*}\d_z G^{*\, -1}
\label{Agauge}
\fin
with $G$ an element of the complex $su(2)^C$ group, i.e. $G$ is a two by two
complex matrix with determinant one. This is always possible on the sphere.
This parametization is such that
$\psi_-(\d_{\bar z} + i \bar{ A}) \psi_+=
(\psi_-G^{-1})\d_{\bar z} (G\psi_+)$. Let us now denote by $\psi'$
the chiral gauge-transformed fermions:
\debut
\psi_-' = \psi_-G^{-1} \quad &;&\quad 
\bar \psi_-' = \bar \psi_-G^{*} \label{chiral}\\
\psi_+' = G\psi_+ \quad &;&\quad
\bar \psi_+' = G^{*\, -1}\bar \psi_+ \non
\fin
Gauge transformed bosons $\beta'$ are defined similarly.
The fermionic action (\ref{2.8}) may be rewritten in terms
of $G$ and $\Psi'$. It becomes:
\debut
\Ga(\Psi'|G,\al,m)&=&
\int \frac{d^2x}{2\pi}\,\Bigl(\, \psi_-'\d_{\bar z} \psi_+' 
+ \bar \psi_-'\d_z \bar \psi_+' \non\\
&~& ~~~~~+\left. i \psi_-' G(\alvec\cdot\sigvec -m)G^*\bar \psi_+ '
+ i\bar \psi_-' G^{*\, -1}( \alvec\cdot\sigvec +m) G^{-1}\psi_+'\)  
\label{actionGm}
\fin
This is the effective action for the Dirac operator coupled to the
disorder variables $G\alvec\cdot\sigvec G^*$, 
$G^{*\, -1}\alvec\cdot\sigvec G^{-1}$ and $G m G^*$.
As we shall see below, the field $G$ will be coupled
to the spin degrees of freedom whereas the fermions $\Psi'$
shall be coupled to the charged sector.
The density of states has a factorized expression
in terms of these new spin and charged variables:
\debut
\rho(E) \propto\ {\rm Re}\ \vev{ \bar \Psi'\, (GG^*) \,\Psi' }
\label{rhofact}
\fin

We must not neglect to take into account 
the jacobians of the transformations $A\to G$ and $\Psi \to \Psi'$
as well as to represent the partition function $Z({A, \al},m)$.
The jacobians, which are determinants of
Dirac operators, are computed using the chiral anomaly. 
They may be expressed with the help of the WZW action:
\debut
\Big\vert \frac{D\Psi}{D\Psi'}\Big\vert &=& {\rm Det}\(i\dsl+A \)
=\exp\(S_{wzw}(GG^*)\,\) \non\\
\Big\vert \frac{DA}{DG}\Big\vert &=&
\exp\(4 S_{wzw}(GG^*)\,\) \non
\fin
with $S_{wzw}$ the WZW action. 

To represent the partition function,  
note that ${\rm Det}\(i\dsl+A\)=Z({ A, \al}=0,m=0)$. Therefore,
chiral gauge transformations applied to the bosonic beta system give:
\debut
\frac{{\rm Det}\(i\dsl+A\)}{Z(A,\al,m)}=
\int D\beta' \exp\(\ -\Ga(\beta'|G, \al, m )\,\)
\label{partition}
\fin
Gathering the jacobians we get the  action 
\debut
S=-4 S_{wzw}(GG^*) + \Ga(\Psi'|G,\al, m)
+\Ga(\beta'|G, \al, m) \label{effaction}
\fin
with $\Ga$ defined in eq.(\ref{actionGm}).
This is the action to compute correlations using $G$, $\Psi'$ and $\beta'$ as 
the path integral variables at fixed disorder. 
One still has to add the disorder measure
(\ref{2.2},\ref{2.4}) to compute averaged correlations.

In eq.(\ref{effaction}) the random spin variables $A$ or $G$ are not yet decoupled.
This decoupling only appears when $g_\al+g_m=0$.
Indeed, in that case integrating over the disorder $\bf \al$ and $m$ yields
to current-current type interactions as in eq.(\ref{4.5}) but 
with $osp(2|2)$ currents bilinear in the fermions $\Psi$ 
instead of the fermions $\Psi'$.
The crucial point is now to remark that these $osp(2|2)$ currents
are invariant under the chiral gauge transformations (\ref{chiral}).
For example,
\debut
\psi_-'\cdot \psi_+' &=& \psi_-\, GG^{-1}\,\psi_+= \psi_-\cdot \psi_+ = J\non\\
\ep_{ij}\, \psi_{+\,i}'\,\psi_{+\,j}'
&=& \psi_{+\,n}\psi_{+\,m}\, G_{in}G_{jm}\,\ep_{ij} 
=\psi_{+\,n}\psi_{+\,m}\,\ep_{nm}=J^- \non
\fin
where we used that $G_{in}G_{jm}\ep_{ij}={\rm det G}\ \ep_{nm}=\ep_{nm}$
since $G$ has unit determinant. Similarly, it is easily checked that 
all $osp(2|2)$ currents are   
invariant under chiral gauge transformations. Note that this
is true because the spin disorder variables belong to $su(2)$.
This would  not be valid if for example the spin disorder variables were taking 
values in $u(2)$ instead.

Hence, after integrating over the disorder at $g_\al+g_m=0$, the spin 
random variables $G$ decouple from the $\Psi'$ and $\beta'$ system.
This is the spin-charge separation. 
We are thus left with the  action:
\debut
S= -4 S_{wzw}(GG^*) + \Ga(\Psi'|G=1,\al,m) 
+\Ga(\beta'|G=1, \al, m) 
\label{decouple}
\fin
with the same disorder measure for $G$ and $\al,\ m$ as in eq.(\ref{2.2},\ref{2.4})
with $g_A$ arbitrary but $g_\al+g_m=0$. 
The first term describes a WZW theory on the coset space  $su(2)^C/su(2)$
at level $k=-4$. This may be thought of  as the theory `inverse' to the
$su(2)$ WZW theory at level $k=0$ \cite{GadKu}, since the conformal dimensions
of primary field in the two theories have opposite sign.              
The second and third terms describe the $osp(2|2)$ current-current
perturbation of the $\Psi'- \beta'$ system.
As in previous section this $\Psi'- \beta'$ system
may also be described as an $osp(4|4)$ WZW model at level one
and the current-current perturbation only couples to the
sub-sector generated by the $osp(2|2)$ currents.

In the last section we argued that $g_A$ flows to infinity under
RG flow. 
At $g_A=\infty$,  the measure (\ref{2.2}) on $G$ is flat and
only the WZW action $S_{wzw}(GG^*)$ at level $k=-4$ 
remains. This describes the spin sector.
 
If $g_\al=-g_m>0$, the $osp(2|2)$ current perturbation is irrelevant
and we recover the previous description. Recall the original
fermion $\Psi$, from which the density of states is computed,
are related to $G$ and $\Psi'$ by the chiral gauge transformation (\ref{chiral}),
so that the density of states is factorized as in eq.(\ref{rhofact}). 

If $g_\al=-g_m<0$, the current-current perturbation is relevant.
We may then propose that in the infrared the $\Psi'- \beta'$ system
is described by the coset theory $osp(4|4)_{k=1}/osp(2|2)_{k=-2}$, which is equivalent
to an $su(2)$ model at level zero. As a consequence, the fermions $\Psi'$
shall flow in the infrared to fields with scaling dimension opposite to
that of $G$, and the original fermions $\Psi$ shall flow
to fields with zero scaling dimension. This suggests that in this
case the density of states is finite and regular at zero energy.

\section{RG Phase diagram}  

In this section, we return to the general model with $g_\al + g_m\neq 0$, 
and describe the global features of the phase diagram based on the 
one-loop RG equations (\ref{3.10}). 

Our  method consists in extracting the asymptotics of the 
RG trajectories by looking for directions in the coupling constant space
which are preserved by the RG flow. Then, to analyze whether these
asymptotic trajectories are attractive or not, stable or unstable,
we project the RG flow onto the sphere and study the vector field
thus obtained. This will allow us to point out
the special role played by perturbations along a  so-called   
strange direction.

Let us first look for lines in the coupling constant space,
with coordinates $g=(g_A,g_\al,g_m)$, which
are preserved by the RG flow and which pass through the origin. 
These correspond to trajectories which are straight lines
and therefore for which the RG velocity field $\dot g = \beta(g)$
is co-linear to the vector $g$. The equations for these fixed
line trajectories are thus:
\debut
\beta(g) \wedge g =0 
\label{lines}
\fin
The above equation is equivalent to finding solutions of the RG
equations of the form $g^i (t) =  x^i \la(t)$
where $x^i$ are constants independent of the RG time $t= \log l$
and $d\la(t)/dt = \la^2$.  Substituting this into (\ref{3.6}),  
one finds
\beq
\label{xxx}
x^i = - C^i_{jk} x^j x^k 
\eeq
where as before $C^i_{jk}$ are the operator product coefficients. 
With the linear relation $g^i =  x^i \la$ among the couplings, the effective action
contains a single running coupling constant:
\beq
\label{Sx}
S_{\rm  eff} = S_{cft} + \la\,\int  \dx \ \CO_x , ~~~~~~~
\CO_x = \sum_i  x^i \CO_i 
=  x^A \CO_A+x^\al\CO_\al+x^m\CO_m
\eeq
For $x^i$ a solution to (\ref{xxx}), the operator $\CO_x$ closes on itself
under operator product expansion:
\beq
\label{xxO}
\CO_x (z, \zbar) \CO_x (0) \sim - \inv{z\zbar} ~ \CO_x (0)
\eeq
To lowest order the $\beta$eta function for this 
specific perturbation is 
$\beta_\la = \la^2 +\cdots$

There are six solutions to eqs.(\ref{lines}) (or eqs. (\ref{xxx})). They are:
\debut
A &\equiv & \quad \(g_A\not=0,\ g_\al=0,\  g_m=0 \)\label{solA}\\
B &\equiv & \quad \(g_A=0,\ g_\al=0,\ g_m \not=0\)\label{solB}\\
C &\equiv & \quad \(g_A=0, g_\al+g_m=0 \)\label{solC}\\
D &\equiv & \quad \(g_\al+g_m=0,\ g_A+2g_\al=0 \)\label{solD}\\
E &\equiv & \quad \(6\,g_\al=-3(1+\sqrt{2})\, g_A = -2(1+2\sqrt{2})\, g_m \)\label{solE}\\
F &\equiv & \quad \(6\,g_\al=3(\sqrt{2} - 1)\, g_A = 2(2\sqrt{2} -1 )\, g_m \)\label{solF}
\fin
The two first lines correspond to the simple models with only
random gauge potential \cite{tsvelik,carg} or only random mass \cite{mass1,mass2,mass3}. 
The third and fourth ones
correspond to the case we discuss at length in section IV.
Contrary to the first four solutions, the last two
do not seem to have an obvious simple algebraic interpretation
\footnote{The operator $\CO_x$ is bilinear in the currents,
$\CO_x=J^Ad_{AB}\bar J^B$ with $J^A$ the $osp(4|4)$ currents and $d_{AB}$
some bilinear form. The condition (\ref{xxO}) is then:
$f^{AB}_I d_{AN}d_{BM}f^{NM}_K=d_{IK}$. The algebraic interpretation
of that equation is not clear to us.}.

These six solutions correspond to twelve possible stable directions
because we have to choose an orientation on the line.
Only three of them are in the domain of positive coupling constants
$g_A>0$, $g_\al>0$ and $g_m>0$. The first two, which are the
solution $A$ with $g_A>0$ and the solution $B$ with $g_m>0$ are in the border
of that domain. The third one, which is the solution $F$ with $g_\al>0$,
sits in the middle of the domain of positive $g$'s. 
We shall call this solution the strange direction.

To analyze deeper the RG flow let us now project it onto the sphere. 
This is worth doing since the beta functions are homogeneous.
Since there are three coupling constants, we may parameterize them 
with two angles, that we shall denote $\th_1$ and $\th_2$, and
the radial coordinates $\rho$, $\rho^2=g_A^2+g_\al^2+g_m^2$.
The RG equations (\ref{3.10}) may then be written as two
equations for the angular variables
\debut
\dot \th_j = \rho \ \beta_j(\th_1,\th_2) \label{betangle}
\fin
together with one equation for the radial variable:
\debut
\dot \rho = \rho^2\ \beta_\rho(\th_1,\th_2) \non
\fin
The explicit expressions for the vector fields $\beta_j$ 
or $\beta_\rho$ are easy to find.
In the angular equations (\ref{betangle}) we may absorb the
factor $\rho$ into a redefinition of the parametrization
of the RG trajectories. This does not change the topology
of the RG curves but only the speed at which the RG trajectories
flow on these curves. So we shall analyze the vector field
$\dot \th_j = \beta_j(\th_1,\th_2)$ on the sphere.
See figure 1.

The vector field $\beta_j$ has twelve zeroes 
which correspond to the twelve fixed directions (\ref{solA}-\ref{solF}).  
We can compute the sign of $\dot \rho$ in these  directions
to know whether these straight RG trajectories are escaping to
infinity ($\dot \rho >0$) or are flowing back to
the origin ($\dot \rho <0$). 
To decipher whether the fixed directions are attractive or not
we have to analyze whether the corresponding zero of the 
vector field $\beta_j$ on the sphere is attractive or not.
For that we linearize the vector field $\beta_j$ on the sphere at its zeroes
and compute its eigenvalues. 
Positive eigenvalues correspond to repulsive fixed directions,
zero eigenvalues to locally marginal directions.
For a fixed direction to be generically attractive, the two
eigenvalues have to be non positive; otherwise to be attracted to
the fixed direction requires fine tuning. 
The result is:


\debut
A_+ (g_A>0)\quad &\equiv & \quad \dot \rho >0 
\quad {\rm and}\quad  {\rm repulsive} \non\\
A_- (g_A<0)\quad &\equiv & \quad \dot \rho <0 
\quad {\rm and}\quad  {\rm repulsive}\non\\
B_- (g_m>0)\quad &\equiv & \quad  \dot \rho <0 
\quad {\rm and}\quad  {\rm repulsive}\non\\
B_+ (g_m<0)\quad &\equiv & \quad \dot \rho >0 
\quad {\rm and}\quad  {\rm attractive}\non\\
C_- (g_\al>0)\quad &\equiv & \quad \dot \rho <0 
\quad {\rm and}\quad  {\rm repulsive}\non\\
C_+ (g_\al<0)\quad &\equiv & \quad \dot \rho >0  
\quad {\rm and}\quad  {\rm attractive}\non\\
D_- (g_\al>0)\quad &\equiv & \quad  \dot \rho <0 
\quad {\rm and}\quad  {\rm attractive}\non\\
D_+ (g_\al<0)\quad &\equiv & \quad  \dot \rho >0 
\quad {\rm and}\quad  {\rm repulsive}\non\\
E_- (g_\al>0)\quad &\equiv & \quad \dot \rho <0 
\quad {\rm and}\quad   {\rm repulsive}\non\\
E_+ (g_\al<0)\quad &\equiv & \quad \dot \rho >0 
\quad {\rm and}\quad  {\rm repulsive}\non\\
F_+ (g_\al>0)\quad &\equiv &\quad  \dot \rho >0 
\quad {\rm and}\quad  {\rm attractive}\non\\
F_- (g_\al<0)\quad &\equiv & \quad \dot \rho <0 
\quad {\rm and}\quad   {\rm repulsive}\non
\fin
The indices $\pm$ refer to trajectories flowing to infinity
or back to the origin.

There are only four fixed directions which are generically attractive:
they stand on the directions $B_+$, $C_+$, $D_-$ and
on the strange direction $F_+$.
They all are asymptotes to RG trajectories flowing to infinity
except $D_-$ which corresponds to asymptotic direction of trajectories
looping back to the origin.

Let us pause   to reconsider the case $g_\al+g_m=0$ is this
language. In that case the solutions of the RG equations are
$1/g_A-1/g_A^0=-2t$ and $1/g_\al-1/g_\al^0=4t$. 
 If $g_A^0>0$ and $g_\al^0>0$, then
$g_A$ increases whereas $g_\al$ decreases. These
trajectories are asymptotic to the direction $A$ since $g_A$ blows up
at a finite value of $t$ at which $g_\al$ is finite.  However the
blow up times are not physical since for
$g_A$ large enough the one-loop analysis is no more valid and one has
to relies on the non-perturbative analysis done in previous section.
If $g_A^0>0$ but $g_\al^0<0$, then both couplings increase in magnitude.
Which one blows up first depends on the initial data.
The asymptotic trajectories are then either the direction $A$ or $C$
depending if $g_A^0+2g_\al^0$ is positive or negative, so that the line
$g_A+2g_\al=0$ is a separatrix in the phase diagram.
On contrary, if $g_A^0<0$ but $g_\al^0>0$, the couplings decrease and flow back
towards the origin along the direction $D$.

The strange direction $F_+$ is the asymptotic 
direction for {\it all trajectories} starting initially
with positive coupling constants $g_{\al, m, A}$, except those which are fine tuned
to be in the direction $A$ or $B$. The domain $g_j>0$ is
stable under one-loop RG: no trajectories can escape from it.
The fact that there is one and only one asymptotic direction
in the domain of positive coupling constants reflects the universality
of that behavior: whatever the values of the initial coupling constants,  
the system flows along this direction in the infrared. 
The strange direction is hence
the one which should be relevant to the description of the low energy
behavior  of the $su(2)$ random Dirac operators, e.g. the network model
described in section II.
Furthermore, since   
$\dot \rho>0$ for the strange direction,  it is a strongly coupled system. 

In summary, based on the one-loop RG equations, it appears the generic
network model is in a different universality class than the one
fine-tuned to the line $g_\alpha + g_m = 0$.  However it  certainly remains
a possibility that the higher loop corrections can spoil the above 
analysis.

\section{Discussion}

In summary, we have shown that the network model with the identification
of couplings $g_\al + g_m = 0$  exhibits a spin-charge separation
in the effective disorder averaged theory, and this allows a precise
identification of the infrared fixed point as the coset
$osp(4|4)_1 / su(2)_0$.  This coset conformal field theory
has some novel features in that it possesses an $osp(2|2)_{-2}$
supercurrent algebra symmetry, but due to the non-factorizability
of the Hilbert space, it is not identical to the $osp(2|2)$ theory
defined by the Sugawara construction.      The resulting critical 
exponents agree with the predictions based on  percolation\cite{Read}
and the numerical
simulations of the super spin chain studied in \cite{Senthil}.  Based
on the one-loop renormalization group, we have
argued that the  network model without the constraint $g_\al + g_m = 0$
is in a different universality class.  

How our analysis relates to percolation remains an interesting open
question.  There appear to be two possibilities.  One is that the
1-loop strange direction we described in section V survives to higher
loops and 
indeed corresponds to percolation; our coset theory $osp(4|4)_1/
su(2)_0$ is then the wrong fixed point.  Another possibility is that
the higher loop corrections actually restore the symmetry to  the
line $g_\alpha + g_m =0$, and the coset $osp(4|4)_1 / su(2)_0$ 
is a new description of percolation with $osp(2|2)$ symmetry. 
The latter possibility can
be investigated by comparing the conformal blocks of percolation
with those computed in \cite{tsvel3}.      

It would be interesting to construct
explicitly the continuum field theory corresponding to the super spin chain
along the lines of \cite{Saleur2}. 
It is important to understand the role of the global $\osp$ symmetry of the 
spin chain if this continuum limit indeed corresponds to percolation.

\section{Acknowledgments}

We thank A. Ludwig,  H. Saleur and D. Serban for discussions, and 
A. La Beli\`ere for spirited  encouragement. D.B. was supported 
in part by the CNRS, by the CEA and the European TMR 
contract ERBFMRXCT960012.

\section{Appendix: spin-charge separation of the stress tensor.}

We give here some   details on the proof of eq.(\ref{4.11}) for the
energy momentum tensor.
Recall that the Sugawara construction of the energy momentum tensor 
in terms of the currents $J^a$ is:
\debut
T(z) = \kappa \lim_{w\to z}\(\, J^a(w)C_{ab}J^b(z) - 
k \frac{C_{ab}\de^{ab}}{(z-w)^2} \,\)
\label{tsug}
\fin
with $C_{ab}$ the quadratic Casimir and the normalization constant
$\kappa$ is chosen such that the currents have conformal dimension one. 
In practice $T$ is computed by extracting the regular term 
in product $J^a (z) C_{ab}J^b (0) $.

For the $osp(2|2)$ algebra at level $-2$ this gives:
\debut
T_{osp(2|2)_{-2}}=\inv{8}\(
J^2 - H^2 - \half(J_-J_+ + J_+J_-) + (S_+S_- - S_-S_+) +
(\hat S_-\hat S_+ - \hat S_+\hat S_-) \ \)\non
\fin
The $osp(2|2)$ currents  are given in eq.(\ref{4.2}) in terms of the
$\beta-\psi$ system. This Sugawara energy momentum tensor  can thus
be expressed in terms of these bosonic and fermionic fields.
A simple computation yields:
\debut
T_{osp(2|2)_{-2}}&=&\inv{8}\Bigl(
\d_z\psi_-\psi_++\d_z\psi_+\psi_- 
+ \d_z\beta_-\beta_+-\d_z\beta_+\beta_- 
- 4 (\psi_-\beta_+)(\beta_-\psi_+) \non\\
&&~~~~~+\left. 3(\psi_-\psi_+)^2 - (\beta_-\beta_+)^2 
+ 2 (\psi_-\psi_+)(\beta_-\beta_+) \)\non
\fin

Similarly, with the normalization (\ref{3.8}) for the
Pauli matrices, the $su(2)$ Sugawara tensor is:
$$
T_{su(2)_0}= \inv{8} \ L^aL^a
$$
with $L^a$ defined in eq.(\ref{4.8}).
Again with the help of the identity (\ref{4.4}), this
may be written in terms of the $\beta-\psi$ system as:
\debut
T_{su(2)_0}&=& \inv{8}\Bigl(
3(\d_z\psi_-\psi_++\d_z\psi_+\psi_- 
+ \d_z\beta_-\beta_+-\d_z\beta_+\beta_-)
+ 4 (\psi_-\beta_+)(\beta_-\psi_+) \non\\
&&~~~~~  - \left. 3(\psi_-\psi_+)^2 + (\beta_-\beta_+)^2 
- 2 (\psi_-\psi_+)(\beta_-\beta_+) \)\non
\fin

Adding both pieces we get:
\debut
T_{osp(2|2)_{-2}}+ T_{su(2)_0}=
\half\(\d_z\psi_-\psi_++\d_z\psi_+\psi_- 
+ \d_z\beta_-\beta_+-\d_z\beta_+\beta_- \) \non
\fin
This is    the energy momentum tensor $T$
for the conformal field theory $S_{cft}$.

\begin{figure}
\caption[]{\label{fig1} 
 The vector field $\beta_j $ on the sphere. We choose the coordinates
$4g_\al=\cos q$, $2(g_\al+g_m)=\sin q \cos p$ and $2g_A=\sin q \sin p$
with $q\in[0,\pi]$ and $p\in[0,2\pi]$.}
\vskip 2.0 truecm
$$\epsfbox{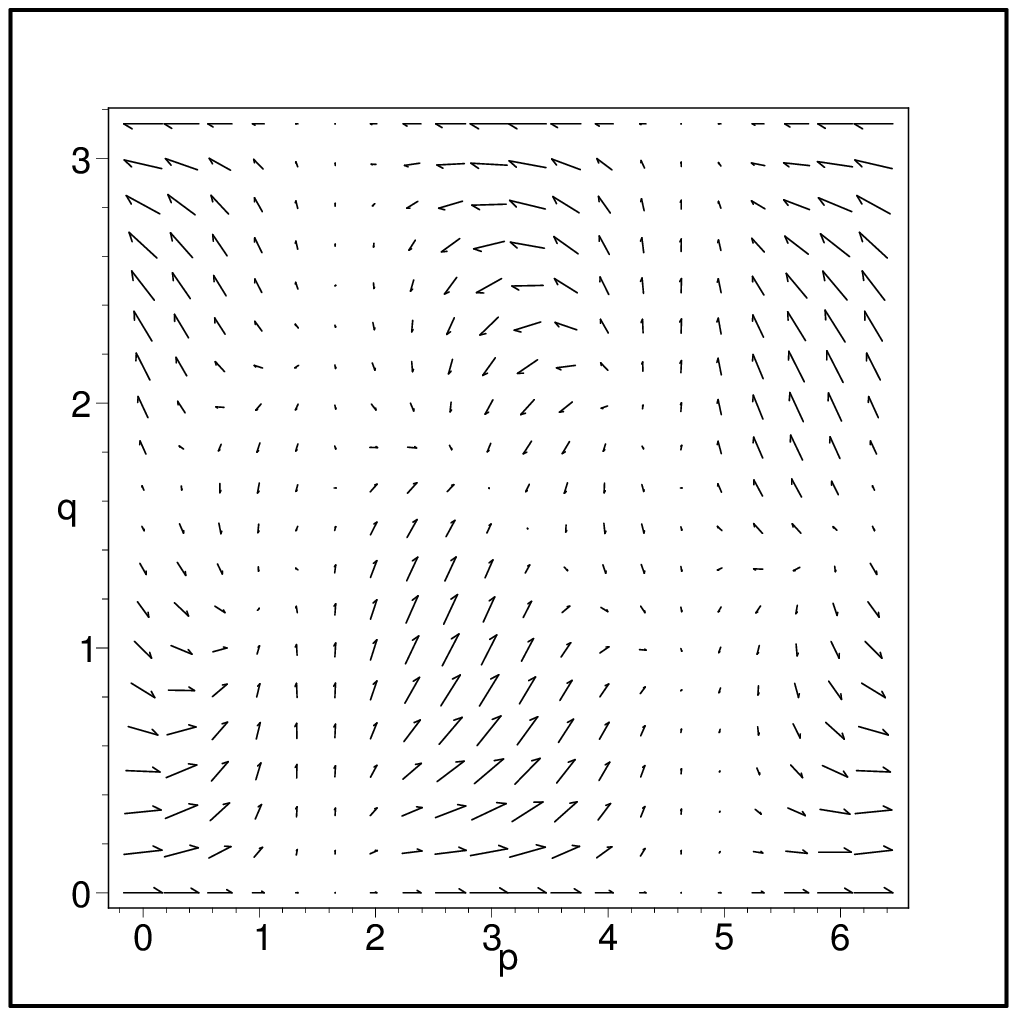}$$


\caption[]{\label{fig2} 
Structure of the indecomposable $osp(2|2)$ representation 
formed by the field appearing in $\tilde \CO$,
$L_S=L_f+L_b= L $ and $L_A=L_f-L_b$. The solid arrows
describe the action of $S_-$ and the dashed ones 
the action of $\hat S_-$.}
\vskip 1.0 truecm
$$\epsfbox{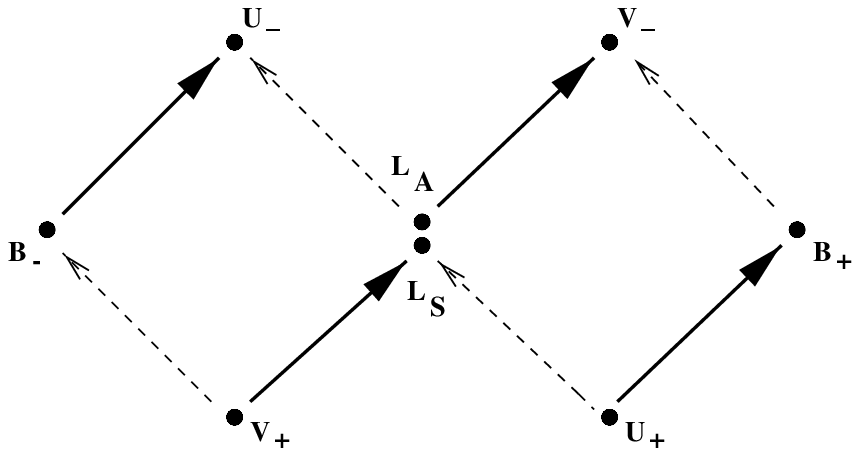}$$

\end{figure}  

\end{document}